\documentclass[twocolumn,showpacs,amsmath,amssymb]{revtex4}
\usepackage{epsfig }
\usepackage{amsmath}
\usepackage{amssymb}

\DeclareMathAlphabet{\bi}{OML}{cmm}{b}{it}

\begin{document}
\title{Landau levels and magnetopolaron effect in dilute GaAs:N}

\author{P. M. Krstaji\'c}

\address{Departement Fysica, Universiteit Antwerpen, \\
Groenenborgerlaan 171, B-2020 Antwerpen, Belgium}
\address{Concordia University, 
7141 Sherbrooke Ouest,
Montr\'{e}al, Canada.}
\author{F. M. Peeters}

\address{Departement Fysica, Universiteit Antwerpen, \\
Groenenborgerlaan 171, B-2020 Antwerpen, Belgium}

\author{M. Helm}

\address{Institute of Ion Beam Physics and Materials Research, \\
Forschungszentrum Dresden-Rossendorf, \\ 
P.O. Box 510119, 01314 Dresden, Germany}

\begin{abstract}
The magnetic-field dependence of the energy spectrum of GaAs doped with nitrogen 
impurities is investigated. Our theoretical model is based on the phenomenological Band Anticrossing 
Model (BAC) which we extended in order to include magnetic field and electron - phonon interaction. Due to 
the highly localized nature of the nitrogen state, we find that the energy levels are very different from those of pure GaAs. 
The polaron correction results in a lower cyclotron resonance energy as compared to pure GaAs. The magneto-absorption spectrum exhibits series of asymmetric peaks 
close to the cyclotron energy $\hbar\omega_c$.
\end{abstract}
\maketitle

\section{Introduction}\label{intro}

Contemporary epitaxial growth techniques provide the possibility for synthesis of high quality semiconductor alloys and/or
elemental materials. Recently, there has been growing interest in N doped GaAs, due to its possible applications for long wavelength 
optoelectronic devices\cite{Kondow}. In GaAs$_{1-x}$N$_x$ substitutional nitrogen has 
the same valence state as As, thus forming an isoelectronic impurity. Experimental data have unambiguously shown that the effects of 
nitrogen incorporation are at least three-fold: a) reduction of the fundamental band-gap\cite{Perkins,Allison,Uesugi} ; b) change in the electron effective mass\cite{Skier,Masia}; and 
c) decrease in the electron mobility\cite{Kurtz}.

Substitutional nitrogen in GaAs forms a resonant level, above the conduction minimum\cite{Wu} ($E_L=0.23{\rm eV}$), having A$_1$ (spherical)
symmetry. Since N(${\rm 2s^22p^3}$) has the same valence state as As(${\rm 4s^24p^3}$) they differ mainly in their local pseudo-potentials resulting in interaction that is 
predominantly short-range\cite{Shan,Lindsay}. It is found that the position of this resonant level,
$E_L$, does not change for nitrogen concentration up to $x=3\%$, and the bowing of the conduction minimum follows a simple square root like law for small $x$. 
Therefore one may expect that a phenomenological approach using perturbation theory is sufficient to explain, for instance, the reduction of
the band-gap. However, this is true only in part, as the significant difference in bond lengths between Ga-As and Ga-N affects next-nearest neighbours\cite{Kleiman}, so that 
the impurity potential has a part which is of intermediate range. What is more, for higher concentration of nitrogen ($x>1\%$), GaAs:N is 
classified as a semiconductor alloy, when the use of Virtual Crystal Approximation (VCA) is questionable, and partial collapse of the Brillouin zone is expected. 
Some authors\cite{Fowler,Endicott} distinguish three ranges for the molar concentration: 1) ultradilute ($x< 0.01\%$), 2) dilute ($0.01\%< x<1\%$), 3) semiconductor alloy ($x>1\%$). 
In this paper we will confine our treatment of GaAs doped with moderately low concentration of N ($x=0.08\%$) which corresponds to case 2). We will investigate the effect of a high 
magnetic field and of electron-phonon interaction on the energy spectrum and on the magneto-absorption spectrum of GaAs$_{1_x}$N$_x$ that has not been considered up to now. 

The paper is organized as follows. In Sect.~\ref{sec_theory} we present our theoretical formalism which is based on the phenomenological Band Anticrossing Model (BAC). 
It is shown how to derive the energy spectrum in the presence of magnetic field in bulk GaAs:N. In the following section, Sect.~\ref{pol} we discuss the 
influence of the electron-phonon interaction on the fundamental transition energy, i.e. the difference between the first two Landau levels. 
In Sect.~\ref{absorsp}, theoretical estimates of the absorption spectrum are given for two different values of temperature and magnetic field. In the last section, Sect.~\ref{conc}, we summarize the results and present our conclusions. 

\section{Theoretical formalism}\label{sec_theory}

The Band Anticrossing Model (BAC) will be employed to determine the energy levels in GaAs:N under an applied magnetic field $B$ with inclusion of electron-phonon
interaction. Within this model the interaction between the nitrogen state and the conduction states is characterized by a single value $C_N$, 
to be defined later. This is possible in dilute GaAs:N where the overlap of neighboring nitrogen wavefunctions is negligible. Then the total wavefunction can
be written as a linear combination of extended states $\Psi^{(0)}_C$ of pure GaAs, and the impurity wavefunction $\Psi^{(0)}_{L}$,

\begin{equation}\label{wave1}
\Psi({\mathbf r})=\alpha\Psi^{(0)}_C({\mathbf r})+\beta\Psi_{L}^{(0)}({\mathbf r})\,.
\end{equation}

It is convenient to represent the conduction band wavefunction $\Psi_C^{(0)}({\mathbf r})$ in the basis of Wannier wavefunctions, $a_C(\mathbf{r-R}_j)$, centered around the sites of the crystal ${\mathbf R}_j$

\begin{equation}\label{wave2}
\Psi_C^{(0)}({\mathbf r})=\frac{1}{\sqrt{M}}\sum_j a_C(\mathbf{r-R}_j)e^{i\mathbf{k\cdot R}_j}\,,
\end{equation}

In the above expression $M$ is the number of Ga-As pairs in arbitrary large volume of the crystal. The matrix element between the nitrogen state and conduction states
of pure GaAs is

\begin{equation}\label{wave3}
\langle\Psi^{(0)}_{L}|\Delta V(r)|\Psi^{(0)}_C\rangle=\sqrt{x}C_N\,,
\end{equation}
where $x$ is the concentration of nitrogen impurities in the semiconductor. Eqs. (\ref{wave1})-(\ref{wave3}) give rise 
to a two-level like secular equation\cite{Shan},
\begin{equation}\label{secular}
\left| \begin{array}{cc}
E-E_C(k) & \sqrt{x}C_N \\
\sqrt{x}C_N & E-E_L
\end{array} \right|=0\,,
\end{equation}
where $E_C(k)$ is the energy dispersion of the conduction band, and $E_L$ is the N-impurity level. The solutions of Eq.~(\ref{secular}) are

\begin{equation}
E_{\pm}=\frac{1}{2}\left\{E_C+E_L\pm\sqrt{(E_C-E_L)^2+4xC^2_{N}}\right\}\,.
\end{equation}

The value of $C_N$ is estimated to be around $2.7{\rm eV}$ by fitting to experimental data\cite{Shan}. The envelope wave function of a nitrogen impurity can be approximated by a Gaussian-like function having only one parameter, the localization radius $a$,

\begin{equation}
\Psi_L({\mathbf r})=\frac{1}{a^{3/2}\pi^{3/4}}e^{-({\mathbf{r-r}}_i)^2/2a^2}\,.
\end{equation}

In the presence of a magnetic field $B$, the conduction band splits in a series of Landau-like levels whose energy spectrum is modified by the presence 
of the short range impurity potential $\Delta V({\mathbf r})$,

\begin{equation}\label{disp}
E_{n\pm}=\frac{1}{2}\left\{ E_{Cn}+E_L\pm\sqrt{(E_{Cn}-E_L)^2+4xC_N^2}\right\}\,.
\end{equation}
The notation $E_{Cn}$ pertains to the pure Landau levels $E_{Cn}=(n+\frac{1}{2})\hbar\omega_c+\hbar^2k_z^2/(2m^{*})$. From now on, index $n$ will refer to pure Landau levels, 
while $n_1$ and $n_2$ will refer to lower and upper subband branches ($n-$ and $n+$ in Eq.~(\ref{disp})). In the absence of impurities, the conduction 
(extended) wavefunctions have the form of a linear harmonic oscillator, and for the magnetic field orientation along the 
$z$ axis and choosing the Landau gauge $A=(-By,0,0)$, their explicit form is as follows

\begin{equation}
\psi_n^{LL}({\mathbf r})=N_n e^{-\frac{(y-y_0)^2}{(2l_c^2)}}H_n\left(\frac{y-y_0}{l_c}\right)e^{i(k_xx+k_zz)}\,.
\end{equation}
Notation $H_n$ corresponds to the $n^{th}$ order Hermitian polynomial, and $y_0$ is the $y-$ coordinate of the center of the orbit, while the normalization constant depends on 
the cyclotron orbit $l_c$, and is given by $N_n=(\sqrt{\pi}l_c2^nn!)^{-1/2}$.

The final wavefunction for the $n_1^{th}$ lower state has the form:
\begin{equation}\label{totalwf}
\Psi_{n_1}({\mathbf r})=\alpha_{n_1}\Psi_{n_1}^{LL}({\mathbf r})+\beta_{n_1}\Psi_L({\mathbf r})\,.
\end{equation}

\begin{figure}[h]
\includegraphics[width=7cm]{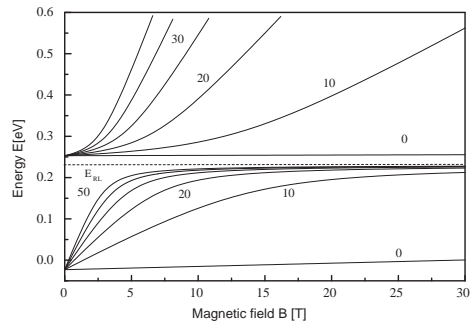}
\caption{\label{fig1} Energy levels in GaAs:N with nitrogen concentration of $x=0.08\%$, as a function of magnetic field $B$, neglecting the interaction 
with phonons. The levels are derived from Landau levels, whose indices are shown near the different curves. It is obvious that the spectrum consists of two 
branches, one of which is below the nitrogen level, $E_L$, while the another one is above.}
\end{figure}

Fig.~\ref{fig1} shows the energy levels in GaAs:N with nitrogen content of $x=0.08\%$, vs magnetic field strength. The interaction with optical phonons is 
neglected for the moment to emphasize the influence of the magnetic field alone. The spectrum splits into two parts with respect to the isolated, localized level 
$E_L$. The lower branch of the levels pin to the value of $E_L$ for high values of magnetic field $B$. At this point it would be useful to inspect the behavior of the 
two branches, $E_{n_1}$ and $E_{n_2}$, Eq.~(\ref{disp}) for small and large values of magnetic field. For small magnetic fields one has the following asymptotic expressions 
\begin{subequations}
\begin{eqnarray}\label{smallBn1}
E_{n_1}=\frac{1}{2}(E_L-D)+A_1\left(n_1+\frac{1}{2}\right)\hbar\omega_c \nonumber \\ 
+B_1\left(n_1+\frac{1}{2}\right)^2(\hbar\omega_c)^2+O((\hbar\omega_c)^3)\,,
\end{eqnarray}
\begin{eqnarray}\label{smallBn2}
E_{n_2}=\frac{1}{2}(E_L+D)+A_2\left(n_2+\frac{1}{2}\right)\hbar\omega_c \nonumber \\ 
+B_2\left(n_2+\frac{1}{2}\right)^2(\hbar\omega_c)^2+O((\hbar\omega_c)^3)\,,
\end{eqnarray}
\end{subequations}
where $D=\sqrt{E_L^2+4xC_N^2}$ and the coefficients $A_{1,2}$ and $B_{1,2}$ are given by the expressions
\begin{eqnarray}
A_1=\frac{1}{2}\left(1+\frac{E_L}{D}\right)\,,\,\,B_1=\frac{1}{2}\left(\frac{E^2_{L}}{D^3}-\frac{1}{D}\right)\,,\,\, \nonumber \\
A_2=\frac{1}{2}\left(1-\frac{E_L}{D}\right)\,,\,\,B_2=-B_1\,.
\end{eqnarray}
The reduction of the bandgap is manifested in Eq.~(\ref{smallBn1}), as the zeroth order term, and is equal to $(E_L-D)/2$. The numerical values of $A_{1,2}$, $B_{1,2}$ for 
the concentration of nitrogen impurities $x=0.08\%$ are $A_1=0.92,\,A_2=0.08$ and $B_1=-0.55=-B_2$. On the other hand, we will also need approximate expressions for large $B$,
\begin{subequations}
\begin{equation}
E_{n_1}=E_L-\frac{2xC_N^2}{(n_1+\frac{1}{2})\hbar\omega_c}+O\left(\frac{1}{(\hbar\omega_c)^2}\right)\,,
\end{equation}
\begin{equation}
E_{n_2}=\left(n_2+\frac{1}{2}\right)\hbar\omega_c+\frac{2xC_N^2}{(n_2+\frac{1}{2})\hbar\omega_c}+O\left(\frac{1}{(\hbar\omega_c)^2}\right)\,.
\end{equation}
\end{subequations}

It is obvious that the lower branch $E_{n_2}$ pins to the nitrogen induced level, $E_L$ for large magnetic fields. 
Furthermore, there is a gap between the values for $B\rightarrow0$ and $B\rightarrow\infty$ which is given by
\begin{equation}
\delta_0=\frac{1}{2}\left(\sqrt{E_L^2+4xC_N^2}-E_L\right)\,.
\end{equation}
This is a consequence of the fact that we have a two-level like problem whose energy separation is determined by the matrix element $C_N$, and the 
concentration $x$. For instance, for $x=0.08\%$, its value is $\delta_0=23{\rm meV}$.

Raman measurements\cite{Bachelier} indicate that the localization radius lies in the range $a_{HW}=1.25-1.7{\rm nm}$ ($a_{HW}=a\sqrt{ln2}$), so that the parameter 
$a$ is of order $2{\rm nm}$. For the present model to be valid, one should impose the condition that the cyclotron orbit is at least three times larger than 
the spatial extent of the impurity wavefunction $|\psi_L|^2$. This ensures that the magnetic field does not distort the impurity wavefunction, 
and it yields the upper bound $B_{max}=35{\rm T}$ for the magnetic field. The coefficients $\alpha$ and $\beta$ are found from their ratio (determined by 
the corresponding eigenvalue equation) and the condition that the wavefunction be normalized (the phase factors of $\alpha$ and $\beta$ are taken to be zero):
\begin{equation}
\frac{\alpha_{n_1}}{\beta_{n_1}}=\frac{\sqrt{x}C_N}{E_{n_1}-E_{Cn_1}},\,\,\alpha_{n_1}^2+\beta_{n_1}^2=1\,,
\end{equation}
which results into

\begin{equation}\label{alphacoeff}
\alpha_{n_1}=\frac{\sqrt{x}C_N}{\sqrt{xC_N^2+(E_{n_1}-E_{Cn_1})^2}}\,,
\end{equation}
\begin{equation}
\beta_{n_1}=\frac{E_{n_1}-E_{Cn_1}}{\sqrt{xC_N^2+(E_{n_1}-E_{Cn_1})^2}}\,.
\end{equation}

These expressions are essentially the same as those of Ref.\cite{Duboz}. The overlap between the localized impurity wavefunction and conduction band states is neglected which is justified for $B<B_{max}=35{\rm T}$. 

\section{Electron-phonon interaction in GaAs:N}\label{pol}
In order to obtain more precise values for the energy levels, one must also take into account  
the interaction of electrons with the thermal vibrations of the crystal. In a polar semiconductor like GaAs, electrons interact with 
longitudinal optical (LO) phonons more strongly than with other types of phonons. They may be assumed dispersionless having energy 
$E_{LO}=\hbar\omega_{LO}=36{\rm meV}$ in GaAs \cite{Lindemann}. Bearing in mind that coupling with the LO phonons is weak in 
common semiconductors, Fr\"ohlich\cite{Frohlich} proposed the following form of the Hamiltonian
\begin{equation}
H=H_0+\hbar\omega_{LO}\cdot\sum_{{\mathbf q}}b^{\dagger}_{{\mathbf q}}b_{{\mathbf q}}+H_{e-ph}
\end{equation}
\noindent where the expression for the interaction part with the LO phonons reads

\begin{equation}
H_{e-ph}=\sum_{\mathbf q}V_{\mathbf q}e^{i{\mathbf{qr}}}(b^{\dagger}_{\mathbf q}+b_{-\mathbf q})\,,
\end{equation}
\noindent with
\begin{equation}
V_{q}^2=\frac{4\pi\alpha\hbar(\hbar\omega_{LO})^{3/2}}{(2m^{*})^{1/2}\Omega q^2}\,.
\end{equation}

The strength of the electron-phonon interaction depends on the dimensionless coupling constant $\alpha$ which for GaAs has a small value of\cite{Lindemann} 
$0.068$. The interaction becomes important when the cyclotron energy $\hbar\omega_c$ approaches the energy of 
the longitudinal optical phonon $\hbar\omega_{LO}$. This corresponds to the situation when, for instance, the energy of the unperturbed ground state with one 
real phonon $|n=0,{\bf{1}}_{ph}\rangle$, $E_0^{(0)}=\hbar\omega_{LO}+1/2\hbar\omega_c$
crosses\cite{Lindemann} the first excited Landau state with no phonons $|n=1,{\bf{0}}_{ph}\rangle$, $E_1^{(0)}=(3/2)\hbar\omega_c$. The electron-phonon interaction 
removes the degeneracy at $\omega_c=\omega_{LO}$ and makes the two level anticross. The renormalized values $E_0$, and $E_1$ of the two levels can be found 
using perturbative methods. However, the usual Rayleigh-Schr\"odinger perturbation theory (RSPT) does not give precise values of the excited states for 
large values of the magnetic field. To overcome this problem, an improved Wigner-Brillouin perturbation theory\cite{Lindemann,Peeters,Wendler} (IWBPT) is usually employed to 
determine the pinning values of the renormalized levels. The energy correction for the $n^{th}$ state within this method is given by the expression\cite{Lindemann}

\begin{equation}
\Delta E_n=\sum_{m=0}^{\infty}\sum_{{\mathbf q}}\frac{|M_{nm}({\mathbf q})|^2}{D_{nm}}\,,
\end{equation}

\noindent and the matrix element $|M_{nm}|^2$ has the following form for the first two energy levels

\begin{subequations}
\begin{equation}
|M_{0m}|^2=|\alpha_0\alpha_m|^2V_q^2\frac{e^{-s}}{m!}s^m\,\,(n=0)\,,
\end{equation}
\begin{equation}
|M_{1m}|^2=|\alpha_1\alpha_m|^2V_q^2\frac{e^{-s}}{m!}s^{m-1}(m-s)^2\,\,(n=1)\,,
\end{equation}
\end{subequations}
where $s=(q_{\perp}l_c)^2/2$. According to this perturbation scheme, the denominator $D_{nm}$ depends on the energy 
correction itself\cite{Peeters}, $\Delta E_n$,
\begin{equation}
D_{nm}=E_n-E_m(-q_z)-\hbar\omega_{LO}+\Delta E_n-\Delta E_0\,,
\end{equation}
so that the solution must be sought self-consistently. The correction to the ground state energy $\Delta E_0$ is equal to its counterpart within 
Rayleigh-Schr\"odinger perturbation theory, i.e. $\Delta E_0=\Delta E_0^{RS}$.  
 \begin{figure}
\includegraphics[width=7cm]{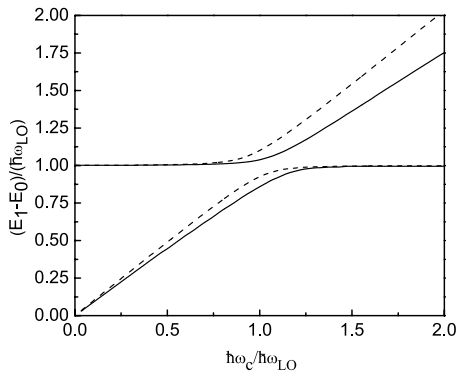}
\caption{\label{fig2} The first transition energy $E_1-E_0$ in GaAs:N as a function of the cyclotron energy $\hbar\omega_c$, in units of 
the LO phonon energy $\hbar\omega_{LO}$. It consists of two branches, one below the phonon continuum and another above it. 
For comparison the values of $E_1-E_0$ for pure GaAs are shown by the dashed curves.}
\end{figure}
Fig.~\ref{fig2} shows the values of the first transition energy $E_1-E_0$ as a function of cyclotron energy $\hbar\omega_c$
(i.e. magnetic field), by the solid curves. For comparison, the values of the same quantity are given for pure GaAs by the dashed curve. The anticrossing behavior of the 
levels $E_0$ and $E_1$ is obvious. The lower curves, which correspond to the case below the phonon continuum pin to $\hbar\omega_{LO}$
for large magnetic fields. The difference between GaAs:N and GaAs is the largest when the cyclotron energy is comparable to the LO phonon energy, $\hbar\omega_{LO}$.
This difference should increase as the cyclotron orbit becomes comparable to the localization radius $l_c\approx a$, but eventually both values should pin to 
$\hbar\omega_{LO}$ for very large $B$.  Another important difference between GaAs:N and pure GaAs is that the values of $E_1-E_0$ above the 
phonon continuum in former (doped) case do not tend to $\hbar\omega_c$, but to a somewhat smaller slope around $A_1$ (see Eq.~(\ref{smallBn1})). This is the consequence of the 
nonparabolicity of the conduction band of GaAs:N. The minimum difference between the two levels is $\Delta E_{10}=5\rm{meV}$ and it occurs at magnetic field $B=23{\rm T}$ just 
above the longitudinal phonon energy $\hbar\omega_{LO}$. 
\section{Cyclotron resonance absorption}\label{absorsp}
Cyclotron resonance measurements is a standard technique\cite{Asch} to measure effective masses in bulk semiconductors. In pure semiconductors, when 
scattering on defects and impurities can be neglected, and for the case of parabolic bands, absorption should ideally consist of a single sharp peak located at the cyclotron 
energy $\hbar\omega_c$. However, in case of GaAs:N, it will be shown 
that the absorption linewidths are naturally broadened due to the change in the conduction band structure caused by the nitrogen impurities. Absorption is a measurable quantity which is 
determined by the oscillator strength that is defined by\cite{Zhang,Jogai,Helm}

\begin{equation}\label{OS1}
I_{fi}=\frac{2}{m^{*}E_{fi}}|\langle i|\hat{p}_y|f\rangle|^2\,.
\end{equation} 
In this work we will consider only transitions between adjacent levels, $n_1-1\rightarrow n_1$, that are normally only possible transitions in cyclotron measurements. In the calculation of the matrix elements $d_{n_1,n_1-1}=\langle n_1-1|{\partial}_y|n_1\rangle$, only one term survives\cite{Duboz} when the total wavefunction, Eq. (\ref{totalwf}) is inserted 
in Eq. (\ref{OS1}), and
\begin{equation}\label{OS2}
d_{n_1,n_1-1}=\frac{\alpha_{n_1}\alpha_{n_1-1}}{l_c}\sqrt{\frac{n_1}2}\,,
\end{equation}
\begin{figure}[h]
\includegraphics[width=7cm]{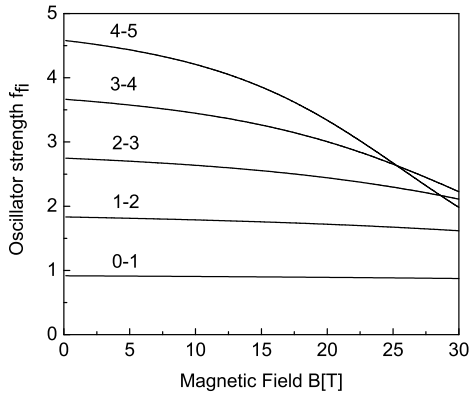}
\caption{\label{fig3} The oscillator strengths for the first 5 transitions between adjacent levels $n_1-1\rightarrow n_1$, below the impurity, localized level $E_L$ as a function of magnetic field $B$.}
\end{figure}

while for the states higher than the impurity state, $E_L$, the expression is the same with $n_1$ replaced by $n_2$. The coefficients $\alpha_{n_1}$ and $\alpha_{n_1-1}$ are given by 
the expression Eq.~(\ref{alphacoeff}).
In the next two figures, Fig.~\ref{fig3} and Fig.~\ref{fig4} we present the oscillator strengths for the first five transitions between adjacent levels $n-1\rightarrow n$, within the lower and upper subbands, 
as a function of magnetic field $B$. 
The values of the oscillator strengths for the lower subbands decrease with increasing magnetic field $B$ (Fig.~\ref{fig3}), due to the presence of coefficients $\alpha_{n_1}$ and $\alpha_{n_1-1}$ in Eq.~(\ref{OS2}). 
This can be explained by the fact that conduction like states $E_{n_1}$ acquire somewhat of a localized nature\cite{Duboz} as they approach the pinning value of $E_L$. On the other hand, the strengths 
for upper subbands start from very small values indicating their highly localized nature at small magnetic fields. 
At the end, it would be useful to calculate the absorption coefficient $\alpha_{abs}$ as a function of the 
energy of the incident light, $\hbar\omega$. Within the dipole approximation, the general formula reads\cite{Seeger}
\begin{figure}
\includegraphics[width=7cm]{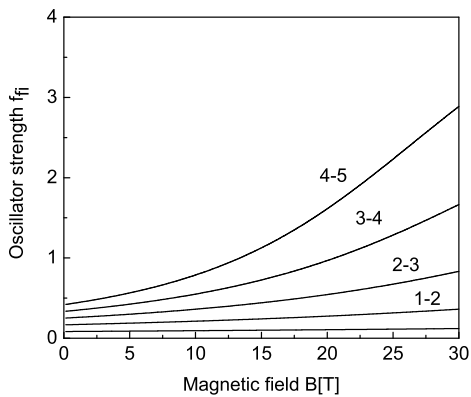}
\caption{\label{fig4} The oscillator strengths for the first 5 transitions between adjacent levels $n_2-1\rightarrow n_2$, above the impurity, localized level $E_L$ as a function of magnetic field $B$.}
\end{figure}

\begin{equation}\label{absor}
\alpha_{abs}(\hbar\omega)=\frac{\mu_0 c}{n_r}\frac{\pi e^2\hbar}{2m^{*}}\sum_{i,f}I_{fi} \rho_j(\hbar\omega)(f(E_i)-f(E_f))
\end{equation} 
\noindent where $n_r$ is the refractive index, $\rho_j$ is the joint density of states, and $\hbar\omega$ is the energy of the incident light. The last factor in Eq.~(\ref{absor}) is the difference in 
the Fermi-Dirac distribution of the initial and the final state ($E_i,\,E_f$).
Note that in Eq.~(\ref{absor}) one should use the modified (joint) density of states, $\rho_j$, to take proper account of the presence of N impurities. The sum in the same equation will in practice terminate  
due to the finite Fermi level and temperature, while in the case of interest $f=i+1$ due to selection rules. In the next figure, Fig.~\ref{fig5} the theoretical estimate 
for the absorption coefficient for intraband transitions within the lower subband are shown as a function of the incident energy $E=\hbar\omega$ at room temperature $T=300{\rm K}$, for two different 
values of magnetic field: $B=10{\rm T}$ (solid curve) and $B=20{\rm T}$ (dashed curve). In both cases the 
electron concentration was kept constant at $n_c=2\times10^{17}{\rm cm^{-3}}$, so that the position of the Fermi level changes with magnetic field. 
To achieve better clarity, a break point is introduced on the $x$-axis between $16$ and $22{\rm meV}$. The broadening and asymmetry of the absorption lines is a consequence of the band nonparabolicity of 
$E_{n_1}(k_z)$ introduced by the nitrogen impurities. For the same reason, the local maxima are not 
located at $\hbar\omega_c$ but do depend on the quantum number $n_1$, since the equation $E_{n_1}(k_z)-E_{n_1-1}(k_z)=\hbar\omega$ is 
not trivial. 
\begin{figure}
\includegraphics[width=7cm]{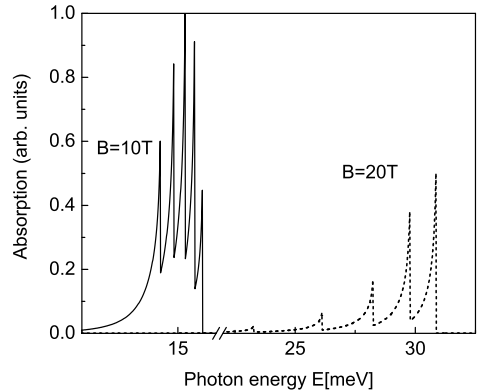}
\caption{\label{fig5} Absorption coefficient $\alpha_{abs}$ in arbitrary units as a function of the incident energy $E=\hbar\omega$ at room temperature $T=300{\rm K}$ for two values of magnetic fields: $B=10{\rm T}$ (solid curve), and 
$B=20{\rm T}$ (dashed curve).}
\end{figure}
\section{Summary and conclusions}\label{conc}
In this work we presented a model to determine the energy levels in dilute nitride GaAs:N under applied magnetic field, taking into 
account electron-phonon interaction. The model is based on the Band Anticrossing Model for GaAs:N but here modified to include 
the interaction with the longitudinal optical phonons treated within second order perturbation theory. It is assumed that the 
impurity wavefunction has a Gaussian shape, spanning just a few lattice constants. The influence of nitrogen
impurities are characterized by three parameters: localized level $E_L$, the matrix element $C_N$ and concentration of the impurities $x$. 
The polaron correction to the energy levels results in a smaller transition energy than from pure GaAs. This difference is the largest 
around the LO phonon energy $\hbar\omega_{LO}$, but eventually becomes zero for large magnetic fields. Furthermore it is shown that the oscillator strength for lower 
subbands (with respect to $E_L$) decrease with increasing of magnetic field as they approach the pinning value of $E_L$. The situation for the upper branch is the opposite, since 
they acquire somewhat of an extended like nature. This should be revealed in the absorption measurement, in the sense that for only higher magnetic fields 
the intraband transition within the upper branch may contribute 
significantly to the absorption spectrum. At the end, the theoretical estimates of the absorption coefficient are done for the lower subband at room temperature $T=300{\rm K}$, for two 
different values of magnetic field $B$. The absorption lines have asymmetric shape due to the non-parabolicity of the conduction band caused by the short range impurity potential.

\section*{Acknowledgments}

This work is supported by the Flemish Science Foundation (FWO-Vl), and the Interuniversity Attraction Poles Program 
(IAP)-Belgian State Science Policy. M.H. is grateful to O. Drachenko and H. Schneider for numerous discussions.


\begin{thebibliography}{00}

\bibitem{Kondow} M. Kondow, T. Kitatani, S. Nakatsuka, M. C. Larson, K. Nakahura, Y. Yazawa, M. Okai, and K. Uomi, IEEE J. Sel. Top. Quantum Electron. 
\textbf{3}, 719 (1997).

\bibitem{Perkins} J. D. Perkins, A. Mascarenhas, Y. Zhang, J. F. Geisz, D. J. Friedman, J. M. Olson, and S. R. Kurtz, Phys. Rev. Lett. \textbf{82}, 3312 (1999).

\bibitem{Allison} G. Allison, N. Mori, A. Patan\`e, J. Endicott, L. Eaves, D. K. Maude, and M. Hopkinson, Phys. Rev. Lett. \textbf{96}, 236802 (2006).

\bibitem{Uesugi} G. Uesugi, N. Marooka and I. Siemune, Appl. Phys. Lett. \textbf{74}, 1254 (1999).

\bibitem{Skier} C. Skierbiszewski, P. Perlin, P. Wisniewski, W. Knap, T. Suski, W. Walukiewicz, W. Shan, K. M. Yu, J. W. Ager, E. E. Haller, J. F. Geisz and J. M. Olson, Appl. Phys. Lett. \textbf{76}, 2409 (2000).

\bibitem{Masia} F. Masia, G. Pettinari, A. Polimeni, M. Felici, A. Miriametro, M. Capizzi, A. Lindsay, S. B. Healy, E. P. OÕReilly, A. Cristofoli, G. Bais, M. Piccin, S. Rubini, F. Martelli, A. Franciosi, P. J. Klar, K. Volz, and W. Stolz
Phys. Rev. B \textbf{73}, 073201 (2006).

\bibitem{Kurtz} S. R. Kurtz, A. A. Allerman, C. H. Seager, R. M. Sieg and E. D. Jones, Appl. Phys. Lett. \textbf{77}, 400 (2000).

\bibitem{Wu} J. Wu, W. Shan, and W. Walukiewicz, Semicond. Sci. Technol. \textbf{17}, 860 (2002).

\bibitem{Shan} W. Shan, W. Walukiewicz, J. W. Ager, E. E. Haller, J. F. Geisz, D. J. Friedman, J. M. Olson, and S. R. Kurtz, Phys. Rev. Lett. \textbf{82}, 1221 (1999).

\bibitem{Lindsay} A. Lindsay and E. P. O'Reailly, Solid State Commun. \textbf{112}, 443 (1999).

\bibitem{Kleiman} G. Kleiman, Phys. Rev. B \textbf{19}, 3198 (1979).

\bibitem{Fowler} D. Fowler, O. Makarovsky, A. Patan\`e, L. Eaves, L. Geelhaar, and H. Riechert,  Phys. Rev. B \textbf{69}, 153305 (2004).

\bibitem{Endicott} J. Endicott,  A. Patan\`e, J. Ib\'a\~nez, L. Eaves, M. Bissiri,  M. Hopkinson, R. Airey, and G. Hill, Phys. Rev. Lett. \textbf{91}, 126802 (2003).

\bibitem{Bachelier} G. Bachelier, A. Mlayah, M. Cazayous, J. Groenen, A. Zwick, H. Carre're, E. Bedel-Pereira, A. Arnoult, A. Rocher, and A. Ponchet, Phys. Rev. B \textbf{67}, 205235 (2003).

\bibitem{Duboz} J.-Y. Duboz, Phys. Rev. B \textbf{75}, 045327 (2007).

\bibitem{Lindemann} G. Lindemann, R. Lassnig, W. Seidenbusch, and E. Gornik, Phys. Rev. B \textbf{28}, 3198 (1983).

\bibitem{Frohlich} H. Fr\"ohlich, Adv. Phys. \textbf{3}, 325 (1954).

\bibitem{Peeters} F. M. Peeters, X. G. Wu, and J. T. Devreese, Phys. Rev. B \textbf{33}, 4338 (1986).

\bibitem{Wendler} L. Wendler, Physica B \textbf{270}, 172 (1999).

\bibitem{Asch} N. W. Ashcroft and N. D. Mermin, \textit{Solid State Physics}, (Saunders College Publishing, Philadelphia, 1976).

\bibitem{Zhang} J.-Z Zhang and I. Galbraith, Phys. Rev. B \textbf{77}, 205319 (2008).

\bibitem{Jogai} B. Jogai and D. N. Talwar, Phys. Rev. B \textbf{54}, 14524 (1996).

\bibitem{Helm} F. M. Peeters, A. Matulis, M. Helm, T. Fromherz, and W. Hilber, Phys. Rev. B \textbf{48}, 12008 (1993).

\bibitem{Seeger} K. Seeger, \textit{Semiconductor Physics}, (Springer-Verlag, Berlin, 1999).

\end{thebibliography}
\end{document}